\documentclass[twocolumn,prl]{revtex4}
\usepackage{float}
\usepackage{makeidx}
\usepackage{graphicx,amsmath}
\usepackage{CJK}

\begin{document}
\begin{CJK*}{GB}{gbsn}

\title{Momentum-resolved Raman spectroscopy of non-interacting ultracold Fermi gas}
\author{ Pengjun Wang, Zhengkun Fu, Lianghui Huang, Jing Zhang$^{\dagger}$}
\affiliation{State Key Laboratory of Quantum Optics and Quantum
Optics Devices, Institute of Opto-Electronics, Shanxi University,
Taiyuan 030006, P. R. China \label{in}}

\begin{abstract}

We report the experiment on probing the one-body spectral function
in a trapped non-interacting $^{40}$K Fermi gas by means of the
momentum-resolved Raman spectroscopy The experimental result is in
good agreement with the expected quadratic dispersion in the
non-interacting regime. Through the comparison with the
radio-frequency spectrum, we found that the Raman spectrum shows
some new characteristics.

\end{abstract}

\maketitle
\end{CJK*}

During the past decade, the remarkable advances in the study of
ultracold atomic gases have promoted the birth to many interesting
research fields. As an interacting quantum system with highly
tunable parameters, it offers us new opportunities to efficiently
simulate quantum condensed matter systems. The observation of
superfluidity of Fermi gases \cite{one1,one2,one3}, the pairing gap
and pseudogap behaviour \cite{one4,one5}, the quantum simulation of
quantum magnetism and antiferromagnetic spin chains in an optical
lattice \cite{one6,one7} and the generation of synthetic gauge
fields of bosons \cite{one9,one11,one11-1,one11-2} and fermions
\cite{one11-3} were considered to be the important milestones in
this field. In the context of the strongly interacting fermionic
atoms at the vicinity of the Feshbach resonance, the crossover from
the Bardeen-Cooper-Schrieffer (BCS) superfluid state to the
Bose-Einstein-condensate (BEC) superfluid state
\cite{one12,one13,one14} has attracted a lot of attentions. Many
tools have been proposed and used to study the strongly interacting
atomic Fermi gases, for example, spatial noise correlations
\cite{one7,one15}, radio-frequency (RF) spectroscopy
\cite{one4,one5,one16,one17,one18,one19,one19-1}, momentum-resolved
stimulated Raman technique \cite{one20,one21} and Bragg spectroscopy
\cite{one22,one23}.

RF spectroscopy technology as a simple and valuable tool has been
used for experimentally studying physical properties of fermionic
ultracold atoms, such as to decide the scattering length near a
Feshbach resonance by directly measuring the RF shift induced by
mean-field energy \cite{one16}, to demonstrate the quantum unitarity
and many body effect \cite{one17}, to probe the occupied density of
single-particle states and energy dispersion through BEC-BCS
crossover \cite{one5}, and to explore the strongly interacting
two-dimensional Fermi gas \cite{one24,one25,one26,one27}. Besides RF
spectroscopy, Raman spectroscopy is also an important tool. In fact,
RF spectroscopy can be regarded as a special case of Raman
spectroscopy with a vanishing transferred momentum. In the
stimulated Raman process, atoms are transferred into a different
internal state by absorbing a photon from a laser beam and
immediately reemitting the absorbed photon into another laser beam
with different frequency and wave vector. Raman spectroscopy has
several significant advantages comparing with RF spectroscopy
\cite{one21}, that are their spatial selectivity to eliminate
inhomogeneous broadening due to the trap potential, tunability of
the transferred momentum from below to well above the Fermi
momentum, and weaker sensitivity to final-state interactions. The
theoretical studies on utilizing Raman spectroscopy to probe the
excitation spectrum of strongly correlated phases of Bose gases
confined to optical lattices \cite{two1,two2}, to detect the energy
structure of bosonic atoms in a 1D lattice \cite{two3} and to
investigate single-particle excitations in normal and superfluid
phases of strong interacting Fermi gases \cite{one20,one21} have
been accomplished. However, the corresponding experimental
demonstration has not be presented so far.

In this paper, we report the first experimental study on the
momentum-resolved Raman spectroscopy of an ultracold non-interacting
Fermi gas, in which the single-particle excitation is probed. The
quadratic dispersion in the non-interacting regime of Fermi gas is
obtained with Raman spectroscopy technique. Comparing with the RF
spectrum, we found the several advantages of Raman spectroscopy.
This detection technology can be easily extended to probe the
characteristics of Fermi gas in the strongly interacting regime.

\begin{figure}
\centerline{
\includegraphics[width=3.5in]{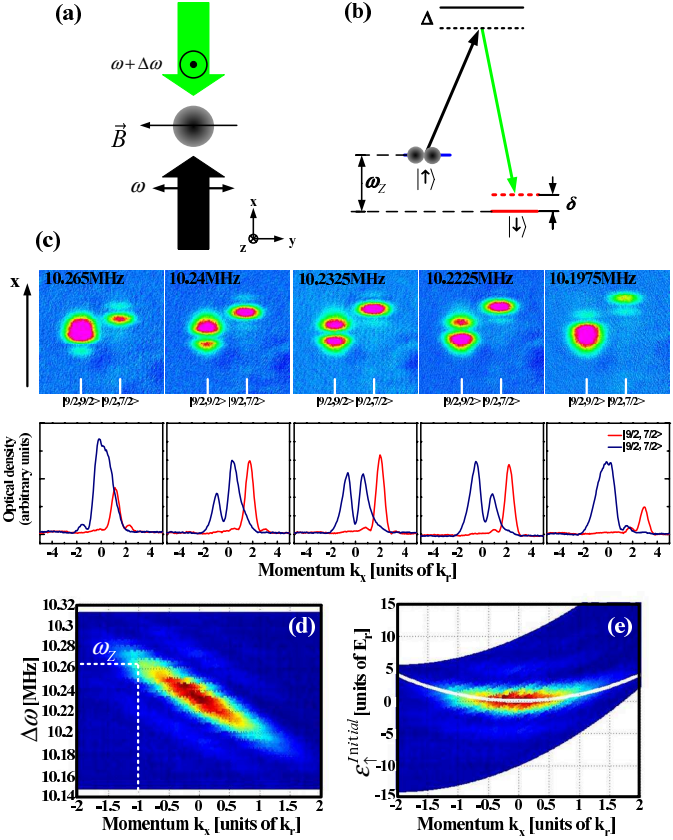}
} \vspace{0.1in}
\caption{(Color online). \textbf{Experiment of Raman spectroscopy}.
\textbf{a,} Schematic of the Raman spectroscopy. The two Raman beams
counterpropagate along $\pm\hat{x}$ with frequency $\omega$ and
($\omega+\Delta\omega$), linearly polarized along $\hat{y}$ and
$\hat{z}$, correspond to $\pi$ and $\sigma$ relative to the
quantization axis $-\hat{y}$. \textbf{b,} The Raman transition with
a Zeeman shift $\omega_{Z}$ and a detuning $\delta$ from the Raman
resonance. \textbf{c,} The absorption images of two hyperfine states
after 12 ms TOF for different Raman frequency detuning. \textbf{d,}
Plot is intensity map of the atoms in the $|\downarrow\rangle$ state
in the ($\Delta\omega, k_{x}$) plane. \textbf{e,} The translated
intensity shows the atom number as a function of the single particle
energy (normalized to $E_{r}$) and momentum $k_{x}$ (normalized to
$k_{r}$). The white line is the expected quadratic dispersion curve.
\label{Fig0} }
\end{figure}

The Raman process is sketched in Fig. 1(a) and (b). The atoms in the
initial hyperfine state $|\uparrow\rangle$ are transferred into the
final empty state $|\downarrow\rangle$ by absorbing a photon from
the laser beam 1 with frequency $\omega_{1}$, wave vector
$\textbf{k}_{1}$, Rabi frequencies $\Omega_{1}$ and then immediately
emitting a photon into another laser beam 2 with frequency
$\omega_{2}$, wave vector $\textbf{k}_{2}$, Rabi frequencies
$\Omega_{1}$ and $\Omega_{2}$. The two laser fields are far detuned
$\Delta$ from resonances with the intermediated excited state so
that the spontaneous emission can be neglected. The effective Raman
coupling of Raman process is defined as
$\Omega=\Omega_{1}\Omega_{2}/\Delta$. When atoms are transt from the
up state to the down state in the Raman process, the momentum
$\textbf{q}_{r}=\textbf{k}_{1}-\textbf{k}_{2}$ and energy
$\hbar\Delta\omega=\hbar(\omega_{2}-\omega_{1})$ are transferred
between photons and atoms. Considering the energy and momentum
conservation in the Raman process, we obtain that
\begin{eqnarray}
\hbar\Delta\omega=E_{Z}+\epsilon_{\uparrow}^{Initial}(\textbf{k})
-\epsilon_{\downarrow}^{Final}(\textbf{k}+\textbf{q}_{r}),
\end{eqnarray}
where $\epsilon_{\uparrow}^{Initial}(\textbf{k})$ and
$\epsilon_{\downarrow}^{Final}(\textbf{k}+\textbf{q}_{r})$ are
energy-momentum dispersion of the spin up and down states
respectively, $E_{Z}$ is the energy split of the two hyperfine
state. From Eq. 1, we can determine the energy-momentum dispersion
of the initial state if the energy-momentum dispersion of the final
state is known (for example,
$\epsilon_{\downarrow}^{Final}(\textbf{k})=\hbar^{2}\textbf{k}^{2}/2m$
for the final state with the non-interaction)
\begin{eqnarray}
\epsilon_{\uparrow}^{Initial}(\textbf{k})=\hbar\Delta\omega-E_{Z}+
\epsilon_{\downarrow}^{Final}(\textbf{k}+\textbf{q}_{r}).
\end{eqnarray}
If we consider the simplest case of the non-interaction fermion gas,
i.e. the energy-momentum dispersions of the initial and final states
present the quadratic function with
$\epsilon_{\uparrow}=\epsilon_{\downarrow}=\hbar^{2}\textbf{k}^{2}/2m$,
Eq. 1 becomes
\begin{eqnarray}
\hbar\Delta\omega=E_{Z}-\frac{\hbar^{2}\textbf{q}^{2}_{r}}{2m}-\frac{\hbar^{2}\textbf{q}_{r}\cdot\textbf{k}}{m}.
\end{eqnarray}
Since the parameters $E_{Z}$ and $\textbf{q}_{r}$ are fixed in
experiment, the function between the frequency difference of two
Raman beams and the atomic momentum is linear.

Next we analyze the RF spectrum. The RF transition is typically
magnetic dipolar transition and the RF spectrum can probe the
single-particle excitation spectrum, in which the momentum of the RF
photon is effectively neglected. We have
\begin{eqnarray}
\hbar\omega_{RF}=E_{Z}+\epsilon_{\uparrow}^{Initial}(\textbf{k})
-\epsilon_{\downarrow}^{Final}(\textbf{k}).
\end{eqnarray}
From Eq. 4, the energy-momentum dispersion of the initial state can
be determined when the energy-momentum dispersion of the final state
is known,
\begin{eqnarray}
\epsilon_{\uparrow}^{Initial}(\textbf{k})=\hbar\omega_{RF}-E_{Z}+
\epsilon_{\downarrow}^{Final}(\textbf{k}).
\end{eqnarray}
Considering non-interacting fermions, the atoms in the hyperfine
states ($|\uparrow\rangle$ and $|\downarrow\rangle$) will experience
the same harmonic trap potential. Since the dispersions of the two
states remain exactly parallel, the RF spectrum will present delta
function $\hbar\omega_{RF}= E_{Z}$.

The experimental setup has been described in our previous works
\cite{three,four,five}. The Bose-Fermi mixtures with $^{87}$Rb at
the spin state $|F=2,m_{F}=2\rangle$ and $^{40}$K atoms at
$|F=9/2,m_{F}=9/2\rangle$ are cooled in magnetic trap and then
transported into an optical dipole trap. The Fermi gas in crossed
optical trap with bosonic $^{87}$Rb further is evaporatively cooled
to $T/T_{F}\approx0.3$  approximately $2\times10^{6}$ $^{40}$K by
reducing the powers of the laser beams, where $T$ is the
temperature, $T_{F}$ is the Fermi temperature defined by
$T_{F}=\frac{\hbar\overline{\omega}}{k_{B}}(6 N)^{1/3}$, and
$\overline{\omega}$ is the trap mean frequency, N is the number of
fermions. When the Fermi gas reaches quantum degeneracy, the optical
trap frequency is $2\pi\times(116,116,164)$ $Hz$ along
($\hat{x}$,$\hat{y}$,$\hat{z}$) for $^{40}$K. In order to remove the
Rb atoms in the trap, we use a Rb resonant laser beam to shine the
mixture in 0.03 $ms$ without heating of $^{40}$K atoms. A pair of
bias magnetic coils are used to create a homogeneous magnetic field
along $\hat{y}$ direction, which generates an energy split between
hyperfine states $|F,m_{F}\rangle=|9/2,9/2\rangle$ and
$|9/2,7/2\rangle$. In the optical trap, two spin states of $^{40}$K
atoms will experience the same trap potential.

For Raman spectroscopy, two $\lambda=773$ $nm$ laser beams
counterpropagating along the $\hat{x}$ axis are linearly polarized
along $\hat{y}$ and $\hat{z}$ respectively, which correspond to
$\pi$ and $\sigma$ relative to quantization axis $\hat{y}$ (as shown
in Fig. 1(a) and (b)). Both beams are extracted from a Ti:sapphire
laser operating at the wavelength of 773 $nm$ with the narrow
linewidth single-frequency and focused onto the central position of
the optical trap with $1/e^{2}$ radii of 200 $\mu m$, which is
larger than the atomic cloud size. Two Raman beams are
frequency-shifted by single-pass through two acousto-optic
modulators (AOM) driven by two signal generators respectively. The
frequency difference of the two Raman lasers $\Delta\omega$ is
adjusted by changing the frequency of the signal generator. We apply
a Raman laser pulse with intensity $I=50$ $mW$ for each laser beam,
and the duration time of 35 $\mu s$, which is much smaller than the
optical trap period. After the Raman pulse, we immediately turn off
the optical trap and the homogeneous magnetic field, let the atoms
ballistically expand in 12 $ms$ and take the time-of-flight (TOF)
absorption image with a CCD (charge-coupled device). In order to
measure the fraction of atoms in different hyperfine states, a
gradient magnetic field along $\hat{y}$ direction is applied with 10
$ms$ during the time-of-flight. The atoms in two spin states are
spatially separated due to the Stern-Gerlach effect. The momentum
transferred to atoms during Raman process is
$|\textbf{q}_{r}|=2k_{r}\sin(\theta)$, where $k_{r}=2\pi /\lambda$
is the single-photon recoil momentum, $\lambda$ is the wavelength of
the Raman beam, and $\theta=180^{o}$ is the intersecting angle of
two Raman beams. Here, $\hbar k_{r}$ and $E_{r}=(\hbar k_{r})^{2}/2m
= h\times 8.34$ $kHz$ are the units of momentum and energy.

All $^{40}$K atoms are initially prepared in the
$|F=9/2,m_{F}=9/2\rangle$ state (spin up state) and the final state
$|F=9/2,m_{F}=7/2\rangle$ is empty. The homogeneous bias magnetic
field is ramped to a certain value, which gives a energy split about
$\omega_{Z}/2\pi\simeq 10.265$ $MHz$ between hyperfine states
$|F,m_{F}\rangle=|9/2,9/2\rangle$ and $|9/2,7/2\rangle$. Then we
apply a Raman pulse with the duration of 35 $\mu s$ to the gas, and
measure the spin population for different frequency differences of
the Raman lasers, as shown in Fig. 1(c). We can see that only atoms
in the certain momentum state are transferred from
$|F=9/2,m_{F}=9/2\rangle$ to $|F=9/2,m_{F}=7/2\rangle$, which is
determined by the frequency difference of the Raman lasers. It
presents the inherent momentum-resolved characteristics of Raman
spectroscopy. We integrate TOF image along $\hat{y}$ to obtain the
momentum distributions in $\hat{x}$ of two spin states respectively,
as shown in Fig. 1(c). The appearance of side lobes in the momentum
distributions of the spin state $|9/2,7/2\rangle$ is due to the
square envelop of Raman laser intensity. Then all momentum
distributions in axis $\hat{x}$ of the spin state $|9/2,7/2\rangle$
for different frequency differences of the Raman lasers are plotted
in the ($\Delta\omega, k_{x}$) plane, as shown in Fig. 1(d), where
all momentum distributions in axis $\hat{x}$ have been translated
for the two-unit momentum -$\textbf{q}_{r}=-2k_{r}$. The
distribution of Fig. 1(d) clearly shows the linear relationship (see
Eq. 3) between the atomic momentum and the frequency difference of
two Raman beams for non-interaction Fermi gas. According to Eq. 2
and the quadratic energy-momentum dispersion of the final state, the
energy-momentum dispersion of the initial state $|9/2,9/2\rangle$
(Fig. 1(e)) is obtained from the measured spectrum (Fig. 1(d)). Fig.
1(e) shows the distribution of the atom number in the initial spin
state as a function of the single particle energy and momentum,
which is in good agreement with the expected behavior of the
quadratic function.

We also carry out the RF spectroscopy in order to compare with the
Raman spectroscopy, which are shown in Fig. 2(a) and (b). A Gaussian
shape pulse of RF is applied to transfer atoms from the initial
state $|9/2,9/2\rangle$ to the final state $|9/2,7/2\rangle$. The RF
radiation is produced by function generator (SRS DS345), and is
controlled by a voltage-controlled RF attenuator for generating
Gaussian shape pulse. The RF pulse is amplified by a power amplifier
(Mini-circuit ZHL-5W-1), then delivered to the atomic cloud by a
simple three-loop coil. The RF Gaussian envelop hence results in the
elimination of the side lobes. With the same homogeneous magnetic
field, we apply a RF Gaussian pulse for 200 $\mu s$ and measure the
spin population for different RF frequencies, as shown in Fig. 2(c).
At the resonance frequency, almost all atoms are transferred to the
$|9/2,7/2\rangle$ state, and no matter how much the momentum of
atoms is. The larger the detuning from the resonance frequency is,
the smaller the number of transferred atoms is. The width of the RF
response is about 3 $kHz$. We integrate TOF image along $\hat{y}$ to
obtain the momentum distributions in $\hat{x}$ of the spin state
$|9/2,7/2\rangle$, then arrange all curves for the different
frequencies of RF pulse into the ($\omega_{RF}, k_{x}$) plane, as
shown in Fig. 2(d). The distribution of Fig. 2(d) shows the delta
function. The energy-momentum dispersion of the initial state
$|9/2,9/2\rangle$ (Fig. 2(e)) can also be obtained from the measured
spectrum (Fig. 2(d)) using to Eq. 5, which is in good agreement with
that measured by the Raman spectroscopy.

\begin{figure}
\centerline{
\includegraphics[width=3.5in]{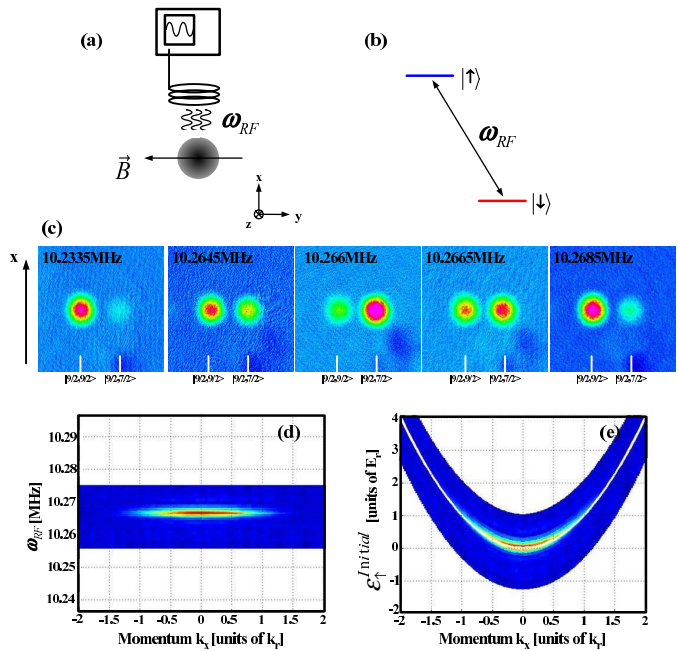}
} \vspace{0.1in}
\caption{(Color online). \textbf{Experiment of RF spectroscopy}.
\textbf{a,} Schematic of the RF spectroscopy experiment. The RF
pulse is coupled to fermionic atoms with a quantization axis
-$\hat{y}$ by a three loop coil. \textbf{b,} Energy level for the
atom-radiation interaction. \textbf{c,} The absorption images of
different hyperfine state after 12 ms TOF for different radio
frequencies. \textbf{d,} Plot is intensity map of the atoms in
$|\downarrow\rangle$ state in the ($\omega_{RF}, k_{x}$)plane.
\textbf{e,} The translated intensity shows the atomic number as a
function of the single particle energy (normalized to $E_{r}$) and
momentum $k_{x}$ (normalized to $k_{r}$). The white line is the
expected quadratic dispersion curve for the non-interacting Fermi
gas. \label{Fig1} }
\end{figure}

In conclusion, we have demonstrated the momentum-resolved Raman
spectroscopy technology experimentally, which is in analogy to the
angle-resolved photoemission spectroscopy of solid state physics.
The single particle property is probed and the dispersion in an
ultracold non-interacting Fermi gas is measured. The experiment
results are good agreement with that obtained with RF spectroscopy
technology. Momentum-resolved Raman spectroscopy technology can be
used to study the single particle state in BEC-BCS crossover, and
the new effects of final states or molecular states, especially if
its spatial selectivity is utilized.

\begin{acknowledgments}

$^{\dagger}$Corresponding author email: jzhang74@yahoo.com,
jzhang74@sxu.edu.cn

This research is supported by National Basic Research Program of
China (Grant No. 2011CB921601), NSFC Project for Excellent Research
Team (Grant No. 61121064), Doctoral Program Foundation of Ministry
of Education China (Grant No. 20111401130001).
\end{acknowledgments}

\end{document}